\documentclass[superscriptaddress,amsmath,amssymb,nofootinbib,eqsecnum,a4paper,secnumarabic,11pt]{revtex4}         
\usepackage[pdftex]{graphicx}
\usepackage{hyperref}     
\usepackage{grffile} 
\newcommand{\VEV}[1]{\left\langle #1 \right\rangle}

\usepackage{titlesec}     
\newcommand*{\justifyheading}{\raggedright}
\titleformat{\chapter}[display]
  {\normalfont\huge\bfseries\justifyheading}{\chaptertitlename\ \thechapter}
  {20pt}{\Huge}
\titleformat{\section}
  {\normalfont\Large\bfseries\justifyheading}{\thesection}{1em}{}
\titleformat{\subsection}
  {\normalfont\large\bfseries\justifyheading}{\thesubsection}{1em}{}
\titleformat{\subsubsection}
  {\normalfont\bfseries\justifyheading}{\thesubsubsection}{1em}{}
\usepackage[english]{babel}

\begin{document}

\title{The effect of the early kinetic decoupling in a fermionic dark matter model}
\author{Tomohiro Abe}
\affiliation{
  Institute for Advanced Research, Nagoya University,
  Furo-cho Chikusa-ku, Nagoya, Aichi, 464-8602 Japan
}
\affiliation{
  Kobayashi-Maskawa Institute for the Origin of Particles and the
  Universe, Nagoya University,
  Furo-cho Chikusa-ku, Nagoya, Aichi, 464-8602 Japan
}

\begin{abstract}
We study the effect of the early kinetic decoupling in a model of fermionic dark matter (DM)
that interacts with the standard model particles only by exchanging the Higgs boson. 
There are two DM-Higgs couplings, namely CP-conserving and CP-violating couplings.
If the mass of the DM is slightly below half of the Higgs boson mass, then
the couplings are suppressed to obtain the measured value of the DM energy density 
by the freeze-out mechanism. 
In addition, the scattering processes of DM off particles in the thermal bath are suppressed by the small momentum transfer if the CP-violating DM-Higgs coupling is larger than the CP-conserving one.
Due to the suppression, the temperature of the DM can differ from the temperature of the thermal bath.
By solving coupled equations for the number density and temperature of the DM,
we calculate the DM-Higgs couplings that reproduce the right amount of the DM relic abundance.
We find that the couplings have to be larger than the one obtained 
without taking into account the difference in the temperatures.
A consequence of the enhancement of the DM-Higgs couplings is the enhancement of the Higgs invisible decay branching ratio.
The enhancement is testable at current and future collider experiments.
\end{abstract}

\maketitle

\section{Introduction}
Dark matter (DM) as a weakly interacting massive particle (WIMP) has been widely studied.
In the WIMP framework, 
DM interacts with the standard model (SM) particles,
and 
pairs of DM particles annihilate into and are created from pairs of SM particles in the thermal plasma in the early Universe. 
These processes are essential for the freeze-out mechanism~\cite{Lee:1977ua} to explain the
measured value of the DM energy density by the thermal relic abundance of WIMP.
The interactions also predict scattering processes of DM particles off nucleons.
A lot of effort has been devoted to detecting such scattering processes directly.
However, DM direct detection experiments did not find significant signals and thus
gave upper bound on the WIMP-nucleon scattering cross sections~\cite{1608.07648, 1708.06917, 1805.12562}.
This upper bound gives a stringent constraint on various WIMP DM models.

One way to avoid the constraint from the direct detection experiments is
to rely on resonance enhancements in the DM annihilation processes.
An example is the Higgs resonance.
If DM interacts with SM particles by exchanging the Higgs boson,
and if the mass of DM is slightly below half of the Higgs boson mass, 
then the annihilation of pairs of DM particles into the SM particles by exchanging
the Higgs boson in $s$-channel is enhanced by the Higgs resonance.
As a result, the DM-Higgs coupling is required to be small to obtain the right amount of the DM relic abundance by the freeze-out mechanism.
On the other hand, the DM-nucleon scattering process is not enhanced by the Higgs resonance because
the scattering is mediated by the Higgs exchange in $t$-channel. 
Therefore, the scattering processes are suppressed by the small coupling,
and thus the models evade the constraints from the direct detection experiments.
This is an excellent feature of the Higgs resonance for DM.
The same mechanism works for other mediator particles if their masses are about twice as large as the mass of DM.

The tiny DM-Higgs coupling makes a difference in temperatures between the dark sector and the visible sector.
In the standard calculation~\cite{Griest:1990kh, Gondolo:1990dk},
it is assumed that the temperatures of DM and the thermal bath are the same as each other; namely, the kinetic equilibrium is assumed.
This assumption is usually relevant in WIMP models because the dark sector and visible sector share the temperature through elastic scatterings of DM particles and particles in the thermal bath.
However, if the elastic scattering is suppressed, then this assumption is not valid,
and the kinetic decoupling can happen earlier than usual.
As a result, the temperature of DM can differ from the temperature of the thermal bath.
It is expected that this early kinetic decoupling happens in the Higgs resonance regime
because the elastic scattering processes are suppressed by the tiny DM-Higgs coupling that is required to obtain the right amount of the DM energy density. 

The authors in \cite{1706.07433} 
developed the method of calculating both the number density and the temperature of DM.
Using their method, 
they studied a scalar singlet DM model~\cite{Silveira:1985rk, McDonald:1993ex, Burgess:2000yq} where a gauge singlet scalar boson is the DM particle and
couples only to the Higgs boson at the renormalizable level. 
They showed that the early kinetic decoupling certainly happens in the Higgs resonance regime.
They also showed that the required DM-Higgs coupling for the thermal relic is bigger than the one in predicted in the standard treatment that ignores the temperature difference between the dark and visible sectors.

In this paper, we apply the method developed in \cite{1706.07433} into an effective theory
of fermionic DM models~\cite{Kanemura:2010sh, 1203.2064, 1205.3169, 1309.3561, Matsumoto:2014rxa, 1512.06458, Matsumoto:2016hbs, 1808.10465}. 
The DM candidate in the model is a gauge singlet Majorana fermion, $\chi$.
It does not couple to the SM fields at the renormalizable level.
Mass dimension-five operators introduce interactions with the Higgs field $(H)$, 
$\bar{\chi} \chi H^\dagger H$ and $\bar{\chi} i \gamma_5\chi H^\dagger H$.
The former respects the CP invariance, while the latter does not.
We focus on the Higgs resonance regime and focus only on these two higher-dimensional operators.
This fermionic DM interacts with the SM particles only through the exchange of the Higgs boson.
The difference between the two types of interactions is important.
For elastic scatterings of DM off SM particles, 
the scattering amplitudes induced by the CP-violating operator are suppressed by the momentum transfer in addition to the small DM-Higgs coupling due to the Higgs resonance.
The momentum transfer is very small
because the DM is non-relativistic in the scattering processes due to the Boltzmann suppression.
Consequently, the scattering is less efficient if the CP-violating operator mainly induces the interaction. 
Therefore, the effect of the early kinetic decoupling is more important in the fermionic DM model with the CP-violating coupling.

The rest of this paper is organized as follows.
In Sec.~\ref{sec:EKD}, we briefly review the early kinetic decoupling.
The zeroth and second moments of the Boltzmann equation are discussed, which have information on the number density and the temperature of DM, respectively.
The coupled equations to be solved are summarized.
In Sec.~\ref{sec:model}, the fermionic DM model is described.
The result with the early kinetic decoupling is discussed in Sec.~\ref{sec:result}.
We show the CP-violating interaction certainly requires larger coupling compared to the one in the standard calculation to obtain the measured value of the DM energy density.
We vary the ratio of the CP-conserving and CP-violating couplings and show that it affects the kinetic decoupling.
Using the values of the couplings required for the right amount of the DM relic abundance,
we discuss the Higgs invisible decay and prospects of its measurements at collider experiments.
We find that the branching ratio of the Higgs decaying into two DM particles can be larger than
the value predicted in the standard calculation.
Section~\ref{sec:summary} is devoted to our conclusion.

\section{The early kinetic decoupling}
\label{sec:EKD}

We briefly review how to calculate the DM number density with taking into account
the effect of the early kinetic decoupling based on the discussion in Ref.~\cite{1706.07433}.

The Boltzmann equation for our universe is given by
\begin{align}
 E 
\left( \frac{\partial}{\partial t} - H \vec{p}\cdot \frac{\partial}{\partial \vec{p}} \right)
f_\chi (t, \vec{p})
=
C_{ann.}[f_\chi] + C_{el.}[f_\chi]
,
\label{eq:boltzmann-eq}
\end{align}
where $E$ is the energy of the DM, $H$ is the Hubble parameter, $\vec{p}$ is the momentum of DM,
and $f_\chi$ is the phase-space density of DM.
The collision term is divided into two parts. 
One is for annihilation of pairs of DM particles ($C_{ann.}$), 
and the other is for elastic scatterings of a DM particle off a SM particle in the thermal bath ($C_{el.}$).
For two-to-two processes, they are written as
\begin{alignat}{3}
 C_{ann.}
=&
\frac{1}{2 g_\chi}
\sum
&
&
\int \frac{d^3 p'}{(2\pi)^3 2 E_{p'}}
\int \frac{d^3 k}{(2\pi)^3 2 E_k}
\int \frac{d^3 k'}{(2\pi)^3 2 E_{k'}}
(2\pi)^4 \delta^4(p + p' - k - k')
\nonumber\\
&&& 
\times \Bigl(
-|{\cal M}_{\chi \chi \to {\cal B} {\cal B}'}|^2 f_\chi(\vec{p}) f_\chi(\vec{p'}) (1 \pm f^{eq}_{\cal B}(\vec{k})) (1 \pm f^{eq}_{\cal B'}(\vec{k'}))
\nonumber\\
&&& \qquad
+|{\cal M}_{{\cal B}{\cal B'} \to \chi \chi}|^2 f^{eq}_{\cal B}(\vec{k}) f^{eq}_{{\cal B}'}(\vec{k'}) (1 \pm f_\chi(\vec{p})) (1 \pm f_\chi(\vec{p'}))
\Bigr)
,\\
 C_{el.}
=&
\frac{1}{2g_\chi}
\sum
&
&
\int \frac{d^3 p'}{(2\pi)^3 2 E_{p'}}
\int \frac{d^3 k}{(2\pi)^3 2 E_k}
\int \frac{d^3 k'}{(2\pi)^3 2 E_{k'}}
(2\pi)^4 \delta^4(p + p' - k - k')
\nonumber\\
&&&
\times \Bigl(
-|{\cal M}_{\chi {\cal B} \to \chi {\cal B}}|^2 
f_\chi(\vec{p}) f^{eq}_{\cal B}(\vec{k}) (1 \pm f_\chi(\vec{p'})) (1 \pm f^{eq.}_{\cal B}(\vec{k'}))
\nonumber\\
&&& \qquad 
+|{\cal M}_{\chi {\cal B} \to \chi {\cal B}}|^2 
f_\chi(\vec{p'}) f^{eq.}_{\cal B}(\vec{k'}) (1 \pm f_\chi(\vec{p})) (1 \pm f^{eq.}_{\cal B}(\vec{k}))
\Bigr)
,
\end{alignat}
where 
${\cal B}$ and ${\cal B}'$ stand for particles in the thermal bath such as quarks, 
$g_\chi$ is the number of internal degrees of freedom of DM,
and
 $f_{\cal B}^{eq}$ is given by the Fermi-Dirac or Bose-Einstein distribution depending on the spin of ${\cal B}$.
The summation should be taken for all the internal degrees of freedom
for all the particles.
For the non-relativistic DM, $C_{el.}$ is simplified as\footnote{
Eq.~\eqref{eq:C_el_apprx} is the same as Eq.~(5) in \cite{1706.07433}.
The expression here makes it clear that $C_{el.}$ does not contribute to the zeroth moment of the Boltzmann equation.
}~\cite{1602.07624}
\begin{align} 
 C_{el.}
\simeq&
 \frac{1}{2 g_\chi} E \frac{\partial}{\partial \vec{p}} \cdot
\Biggl\{
\frac{1}{384\pi^3 m_\chi^3 T}
\int d E_k 
f_{\cal B}^{eq}(E_k)
(1 \pm f_{\cal B}^{eq}(E_k))
\nonumber\\
& \qquad \qquad \qquad \qquad \qquad \times
\int_{-4 k_{cm}^2}^0 dt
(-t) \sum |{\cal M}_{\chi {\cal B} \to \chi {\cal B}}|^2 
\left(
m_\chi T \frac{\partial}{\partial \vec{p}} f_\chi + \vec{p} f_\chi
\right)
\Biggr\}
,
\label{eq:C_el_apprx}
\end{align}
where
$k_{cm}^2$ is given by 
\begin{align}
 k_{cm}^2 
=&
\frac{m_\chi^2 ( E_k^2 - m_{\cal B}^2)}{m_\chi^2 + m_{\cal B}^2 + 2 m_\chi E_k}
.
\end{align}
Here $E_k$ is the energy of ${\cal B}$.
Note that $k_{cm}^2 \neq  E_k^2 - m_{\cal B}^2 = |\vec{k}|^2$.

The temperature of the DM, $T_\chi$, and a related variable $y$ are defined by
\begin{align}
 T_\chi
=&
\frac{g_\chi}{3 n_\chi}
\int \frac{d^3 p}{(2\pi)^3} \frac{\vec{p}^2}{E} f_\chi (\vec{p})
=
\frac{s^{2/3}}{m_\chi} y
,
\end{align}
where $n_\chi$ is the number density of the DM,
and $s$ is the entropy density.
Here $s$ is a function of the temperature of the thermal bath, $T$.
From this definition, $T_\chi$ and $y$ are the function of $T$.
The yield and $x$ are defined as usual,
\begin{align}
 Y = \frac{n_\chi}{s},\
 x = \frac{m_\chi}{T}.
\end{align}
Note that $x$ is defined by $T$ not $T_\chi$.
Differential equations for $Y$ and $y$ are obtained from the zeroth and second moments of the Boltzmann equation, namely
$g_\chi \int \frac{d^3 p}{(2\pi)^3} \frac{1}{E} \times \text{Eq.}~\eqref{eq:boltzmann-eq}$
and
$g_\chi \int \frac{d^3 p}{(2\pi)^3} \frac{1}{E} \frac{\vec{p}^2}{E^2} \times \text{Eq.}~\eqref{eq:boltzmann-eq}$.
Note that the elastic scattering term given in Eq.~\eqref{eq:C_el_apprx} does not contribute to 
the zeroth moment term. This is a natural consequence because the elastic scattering processes do not change the number density of DM.
After some algebra, the following coupled equations are obtained.
\begin{align}
 \frac{dY}{dx}
=&
\sqrt{\frac{8 m_{pl}^2 \pi^2}{45}} \frac{m_\chi}{x^2} \sqrt{g_* (T)}
\left(
-\VEV{\sigma v}_{T_\chi} Y^2
+\VEV{\sigma v}_{T} Y_{eq}^2
\right)
\label{eq:dYdx},\\
\frac{1}{y} \frac{dy}{dx}
=&
\sqrt{\frac{8 m_{pl}^2 \pi^2}{45}} \frac{m_\chi}{x^2} \sqrt{g_* (T)}
\Biggl\{
Y \left(
\VEV{\sigma v}_{T_\chi} 
-\VEV{\sigma v}_{2, T_\chi}
\right)
+
\frac{Y_{eq}^2}{Y} \left(
\frac{y_{eq}}{y}
\VEV{\sigma v}_{2, T} 
-\VEV{\sigma v}_{T}
\right)
\Biggr\}
\label{eq:dydx}
\nonumber\\
&
+
\sqrt{g_* (T)} \frac{x^2}{g_s(T)}
\tilde{\gamma}
\left( \frac{y_{eq}}{y} - 1 \right)
+
\left(
1 + \frac{T}{3 g_s(T)} \frac{d g_s(T)}{dT}
\right)
\frac{1}{3 m_\chi} \frac{y_{eq}}{y} 
\VEV{\frac{p^4}{E^3}}
,
\end{align}  
where
\begin{align}
 \sqrt{g_*(T)}
=&
 \frac{g_s(T)}{\sqrt{g(T)}} \left( 1 + \frac{T}{3 g_s(T)} \frac{d g_s(T)}{dT} \right)
,\\
 \tilde{\gamma}
=&
\sqrt{\frac{8 m_{pl}^2 \pi^2}{45}} 
\frac{15}{256 \pi^5 m_\chi^6 g_\chi}
\sum_{\cal B}
\int_{m_{\cal B}}^\infty dE_k
f_{\cal B}^{eq}(E_k) ( 1 \pm f_{\cal B}^{eq}(E_k))
\int_{-4 k_{cm}^2}^0 dt
(-t)
\sum |{\cal M}_{\chi {\cal B} \to \chi {\cal B}}|^2
\nonumber\\
=&
\sqrt{\frac{8 m_{pl}^2 \pi^2}{45}} 
\frac{15 T}{16 \pi^5 m_\chi^6 g_\chi}
\sum_{\cal B}
\int_{m_{\cal B}}^\infty dE_k
f_{\cal B}^{eq}(E_k) 
\frac{\partial k_{cm}^2}{\partial E_k} k_{cm}^2
\left.
\sum |{\cal M}_{\chi {\cal B} \to \chi {\cal B}}|^2
\right|_{t = -4 k_{cm}^2}
\label{eq:def-gamma-tilde}
,\\
\VEV{\sigma v}_{T_\chi}
=&
\frac{g_\chi^2}{(n_\chi^{eq})^2}
\int \frac{d^3 p}{(2\pi)^3} 
\int \frac{d^3 q}{(2\pi)^3} 
\left( \sigma v \right)_{\chi \chi \to {\cal B} {\cal B}'}
f_\chi^{eq}(\vec{p}, T_\chi)
f_\chi^{eq}(\vec{q}, T_\chi)
,\\
\VEV{\sigma v}_{2, T_\chi}
=&
\frac{g_\chi^2}{(n_\chi^{eq})^2 T_\chi}
\int \frac{d^3 p}{(2\pi)^3} 
\int \frac{d^3 q}{(2\pi)^3} 
\frac{\vec{p} \cdot \vec{p} }{3E} 
\left( \sigma v \right)_{\chi \chi \to {\cal B} {\cal B}'}
f_\chi^{eq}(\vec{p}, T_\chi)
f_\chi^{eq}(\vec{q}, T_\chi)
,\\
\VEV{\frac{p^4}{E^3}}
=&
\frac{g_\chi}{n_\chi^{eq}(T_\chi)}
\int \frac{d^3 p}{(2\pi)^3} \frac{\left( \vec{p} \cdot \vec{p} \right)^2}{E^3} e^{- \frac{E}{T_\chi}}
.
\end{align}
Here
$g$ and $g_s$ are the effective degrees of freedom for the energy and entropy densities respectively,
$f_\chi^{eq}$ is given by the Boltzmann distribution,
and
$m_{pl}$ is the reduced Plank mass, $m_{pl} = 1.220910 \times 10^{19} (8\pi)^{-1/2}$~GeV.
For 
$\VEV{\sigma v}_{T} $
and 
$\VEV{\sigma v}_{2, T} $,
replace $T_\chi$ by $T$ in 
$\VEV{\sigma v}_{T_\chi} $
and 
$\VEV{\sigma v}_{2, T_\chi} $, respectively.
$n_\chi^{eq}(T_\chi)$ is given by
\begin{align}
 n_\chi^{eq}(T_\chi) = g_\chi \int \frac{d^3 p}{(2\pi)^3} f_\chi^{eq}(\vec{p}, T_\chi)
= g_\chi \int \frac{d^3 p}{(2\pi)^3} e^{- \frac{E_p}{T_\chi}}.
\end{align} 
From the first to the second line in Eq.~\eqref{eq:def-gamma-tilde},
we used the following relation,
\begin{align}
 f_{\cal B}^{eq}(E_k) ( 1 \pm f_{\cal B}^{eq}(E_k))
=
- T \frac{\partial}{\partial E_k}  f_{\cal B}^{eq}(E_k) 
,
\end{align}
and integration by parts.


During the QCD phase transition, 
we cannot treat particles as free particles.
Dedicated studies are required for that regime.
In Ref.~\cite{1503.03513},
the table is provided for $g_*$ and $g_s$ for $0.036$~MeV $\lesssim T \lesssim 8.6$~TeV.
Since the values of $g_*$ and $g_s$ do not change for $T \lesssim 0.036$~MeV,
we can regard the values of $g_*$ and $g_s$ at $T = 0.036$~MeV as the values at the temperature today.

We solve 
Eqs.~\eqref{eq:dYdx} and \eqref{eq:dydx}
numerically with the following initial condition
\begin{align}
 Y(x_{ini.}) =& Y_{eq}(x_{ini.}),\\
 y(x_{ini.}) =& y_{eq}(x_{ini.}),
\end{align} 
where $x_{ini.} \simeq 10$.
After solving the coupled equations and obtain $Y(x_0)$, where $x_0$ is 
defined by the temperature of the current universe $T_0$ as 
$x_0 = m_\chi/T_0$,
we convert $Y(x_0)$ into $\Omega h^2$ that is given by
\begin{align}
 \Omega h^2
= \frac{m_\chi s_0 Y(x_0)}{\rho_{cr.} h^{-2}}
,
\end{align}
where~\cite{Tanabashi:2018oca}
\begin{align}
 s_0 =& \frac{2 \pi^2}{45} g_s(x_0) T_0^3,\\
 \rho_{cr.} h^{-2} =& 1.05371 \times 10^{-5} \text{ [GeV} \text{ cm}^{-3}],\\
 T_0 =& 2.35 \times 10^{-13} \text{ [GeV]}.
\end{align}
The measured value of $\Omega h^2$ by the Planck Collaboration
is $\Omega h^2 = 0.120\pm 0.001$~\cite{1807.06209}.
We can use this value to determine a model parameter.

\section{Model}\label{sec:model}

We describe a model that we investigate in the following.
We consider a gauge singlet Majorana fermion DM.
A discrete symmetry $Z_2$ is assumed to stabilize the DM particle. 
Under the $Z_2$ symmetry, the DM is odd while all the other particles, namely the SM particles, are even.
Then, renormalizable operators composed of the DM and SM fields are forbidden.
The DM particle interacts with the SM particles through higher-dimensional operators. 
Therefore, the model is regarded as an effective theory of fermionic DM models.
Up to dimension-five operators, the Lagrangian is given by
\begin{align}
 {\cal L}
=&
 {\cal L}_{\text{SM}}
 + \frac{1}{2} \bar{\chi} \left( i \gamma^{\mu} \partial_\mu  - m_{\chi} \right) \chi
 + \frac{c_s}{2} \bar{\chi} \chi \left( H^\dagger H - \frac{v^2}{2}\right)
 + \frac{c_p}{2} \bar{\chi} i \gamma_5 \chi \left( H^\dagger H - \frac{v^2}{2}\right)
,
\label{eq:eff_lag}
\end{align}
where $\chi$ is the DM candidate, $H$ is the SM Higgs field,
and $v$ is the vacuum expectation value of the Higgs field, $v \simeq 246$~GeV.
The three parameters ($m_\chi$, $c_s$, and $c_p$) are real.
There are two dimension-five operators, $\bar{\chi}\chi H^\dagger H$ and
$\bar{\chi} i \gamma_5 \chi H^\dagger H$.
The former is a CP-conserving operator, while the latter violates the CP invariance.
The CP-conserving interaction has been studied in Ref.~\cite{Kanemura:2010sh},
and the CP-violating operator has been studied in Refs.~\cite{1203.2064, 1205.3169, 1309.3561, 1512.06458,1808.10465, 1901.02278}. 
The DM interacts with the SM particles only by exchanging the Higgs boson
under this setup.

We focus on the mass range 50~GeV $< m_\chi < m_h/2 \simeq 62.5$~GeV.
In this mass range, pairs of the DM particles mainly annihilate into $b\bar{b}$.
The amplitude squared of the annihilation process, 
$\chi \chi \to b \bar{b}$
is given by
\begin{align}
\int \frac{d^3p_b}{(2\pi)^3 2 E_{p_{b}}} \int \frac{d^3p_{\bar{b}}}{(2\pi)^3 2 E_{p_{\bar{b}}}}
 \sum \left| {\cal M}_{\chi \chi \to b \bar{b}} \right|^2
=&
\frac{4 v^2 \sqrt{s} \left( c_s^2 ( s - 4 m_\chi^2) + c_p^2 s \right)
}{
(s-m_h^2)^2 + s \Gamma(\sqrt{s})^2
}
\Gamma(\sqrt{s})_{h \to b\bar{b}}
,
\end{align}
where $\Gamma(\sqrt{s})_{h \to b\bar{b}}$ is the partial decay width of the Higgs boson into $b\bar{b}$, and $\Gamma(\sqrt{s})$ in the denominator is the total decay width of the Higgs boson. Here we take summation for all the internal degrees of freedom for both the initial and final states.
Other annihilation processes are also calculated by replacing $\Gamma(\sqrt{s})_{h \to b \bar{b}}$ properly as long as the pairs of DM particles annihilate through the Higgs boson exchange in the $s$-channel.
The exception is $\chi \chi \to h h$, which contains diagrams exchanging $\chi$ in the $t$ and $u$-channels. However, it is kinematically suppressed in the mass range we are focusing and thus negligible.
The total decay width in the denominator is given by
\begin{align}
 \Gamma(\sqrt{s}) =&
\Gamma_{h}^\text{SM}(\sqrt{s})
+
\frac{\sqrt{s} v^2}{8 \pi}
\left(
  c_s^2 \beta_\chi^3
+ c_p^2 \beta_\chi
\right)
\theta( \sqrt{s} - 2 m_\chi)
,
\end{align}
where 
$\Gamma_{h}^\text{SM}(\sqrt{s})$ is the total decay width of the Higgs boson in the SM particles with $m_h = \sqrt{s}$,
and
\begin{align}
 \beta_\chi =&\left( 1 - \frac{4 m_\chi^2}{s} \right)^{1/2}.
\end{align}
 We use the decay width obtained by the LHC Higgs Cross Section Working Group~\cite{1307.1347} for $\Gamma_h^\text{SM}$.\footnote{
The table is given at \url{https://twiki.cern.ch/twiki/pub/LHCPhysics/CERNYellowReportPageAt8TeV2014/Higgs_XSBR_YR3_update.xlsx}
.} 
Using the amplitude squared above, 
$\VEV{\sigma v}_{T_\chi}$ and $\VEV{\sigma v}_{2, T_\chi}$ are given by
\begin{align}
 \VEV{\sigma v}_{T_\chi}
=&
 \frac{v^2}{16 m_\chi^4 [K_2(\frac{m_\chi}{T_\chi})]^2 T_\chi}
\int_{4 m_\chi^2}^\infty ds
K_1 \left( \frac{\sqrt{s}}{T_\chi}  \right)
\frac{s^2 \Gamma_h^\text{SM}(\sqrt{s})}{(s- m_h^2)^2 + s \Gamma(\sqrt{s})^2 }
\left(
  c_s^2 \beta_\chi^3
+ c_p^2 \beta_\chi
\right)
,
\label{eq:sigmav}
\\
 \VEV{\sigma v}_{2,T_\chi}
=&
 \frac{v^2}{12 T_\chi^3 [K_2(\frac{m_\chi}{T_\chi})]^2}
\int_{1}^\infty d\tilde{s}
\frac{\Gamma^\text{SM}(2 m_\chi \tilde{s}^{1/2}) 
\left( c_s^2 ( \tilde{s} - 1) + c_p^2 \tilde{s} \right)
}{\left( \tilde{s} - \frac{m_h^2}{4 m_\chi^2}  \right)^2 + \frac{\tilde{s}}{4 m_\chi^2} \Gamma(2 m_\chi \sqrt{\tilde{s}})^2}
\tilde{s}^{3/2}
g(\tilde{s})
,
\label{eq:sigmav2}
\end{align}
where $K_n$ is the modified Bessel function of the second kind, and
\begin{align}
 g(\tilde{s})
=&
 \frac{\sqrt{\tilde{s} - 1}}{\sqrt{\tilde{s}}} x K_2(2 \frac{m_\chi}{T_\chi} \sqrt{\tilde{s}})
+
\frac{1}{\sqrt{\tilde{s}}}
\int_1^\infty d\epsilon_+
\exp(- 2 x \sqrt{\tilde{s}})
\ln \frac{\epsilon_+ \sqrt{\tilde{s}} - \sqrt{(\tilde{s}-1)(\epsilon_+^2 -1)} }{\epsilon_+ \sqrt{\tilde{s}} + \sqrt{(\tilde{s}-1)(\epsilon_+^2 -1)} }.
\end{align}

The amplitude squared of the DM-fermion elastic scattering processes are given by
\begin{align}
 \sum |{\cal M}_{\chi f \to \chi f}|^2 
=& 
4 m_f^2 N_c^f
\left[
  c_s^2 \frac{(4 m_\chi^2 - t)(4 m_f^2 - t)}{(t - m_h^2)^2}
+ c_p^2 \frac{(-t)(4 m_f^2 - t)}{(t - m_h^2)^2}
\right]
.
\label{eq:amp_DM-f-ela-scatt.}
\end{align}
Using this amplitude, we find
\begin{alignat}{3}
 \tilde{\gamma} 
=&
\sqrt{\frac{8 m_{pl}^2 \pi^2}{45}}
\frac{60 T}{\pi^5 m_\chi^4}
\sum_{f, \bar{f}}
m_f^2 N_c^f
& \int_{m_f}^\infty d E_k f_f^{eq}(E_k)
k_{cm}^2
\frac{E_k (m_\chi^2 + m_f^2 + m_\chi E_k) + m_\chi m_f^2}{(m_\chi^2 + m_f^2 + 2 m_\chi E_k)^2}
\nonumber\\
&& \times
\frac{c_s^2 ( m_\chi^2 + k_{cm}^2)(m_f^2 + k_{cm}^2) + c_p^2 k_{cm}^2 ( m_f^2 + k_{cm}^2)}{(4 k_{cm}^2 + m_h^2)^2}
.
\label{eq:gamma-tilde-F1}
\end{alignat}
Note that the summation runs both for particles and anti-particles separately,
and $T$ is the temperature of the thermal bath, not of the DM.

As can be seen from Eq.~\eqref{eq:amp_DM-f-ela-scatt.}, 
the CP-violating contribution, which is proportional to $c_p^2$, vanishes as $t$ goes to 0, while the CP-conserving contribution does not.
Since the large $t$ contribution is suppressed by the distribution function (see Eqs.~\eqref{eq:def-gamma-tilde} and \eqref{eq:gamma-tilde-F1}),
the DM-fermion elastic scattering processes are suppressed for $|c_s| \ll |c_p|$.
Therefore, the effect of the early kinetic decoupling is significant for $|c_s| \ll |c_p|$.
This point will be discussed quantitatively in the next section.

In the mass range we focus, the freeze-out happens around $T \simeq {\cal O}(1)$~GeV.
This temperature is not far from the temperature of the QCD phase transition.
Hence the scattering rate of DM and quarks in the thermal bath is potentially affected by the details of the QCD phase transition. The dedicated study is beyond the scope of our work.
Following to Ref.~\cite{1706.07433}, we investigate the two extreme scenarios, QCD-A and QCD-B.
\begin{description}
 \item[QCD-A] All quarks are free particles and present in the thermal bath down to $T_c = 154$~MeV~\cite{1205.1914}. 
 \item[QCD-B] Only the light quarks ($u$, $d$, $s$) contribute to the scattering above $4 T_c \sim 600$~MeV~\cite{0903.0189}.
\end{description}
The difference between these two scenarios is whether charm and bottom quarks contribute to the elastic scattering processes or not. Since the scattering rate is proportional to the squared of the Yukawa coupling of the quark and the color factor, the absence of the heavy quarks makes a large difference between these two scenarios. 
The scattering ratio in the QCD-B is smaller than one in the QCD-A.

\section{Result}\label{sec:result}

We investigate the effect of the early kinetic decoupling on the model described in Sec.~\ref{sec:model}.
We solve
Eqs.~\eqref{eq:dYdx} and \eqref{eq:dydx} numerically
and obtain $\Omega h^2$ for a given parameter set of $c_s$, $c_p$, and $m_\chi$.
One of the model parameters is determined to obtain the measured value of the DM energy density,
$\Omega h^2 = 0.120\pm 0.001$~\cite{1807.06209}.

We start by investigating the maximal CP-violating case ($c_s = 0$) 
because the effect of the early kinetic decoupling is most efficient in that case. 
We also discuss how large the effect of the early kinetic decoupling remains with the CP-conserving coupling.
After determining the couplings, 
we investigate the Higgs invisible decay and the DM-nucleon scattering cross section to discuss the impact of the early kinetic decoupling on phenomenology.

\subsection{Maximal CP-violating case}
\label{sec:CPV}
\begin{figure}[tb]
\includegraphics[width=0.48\hsize]{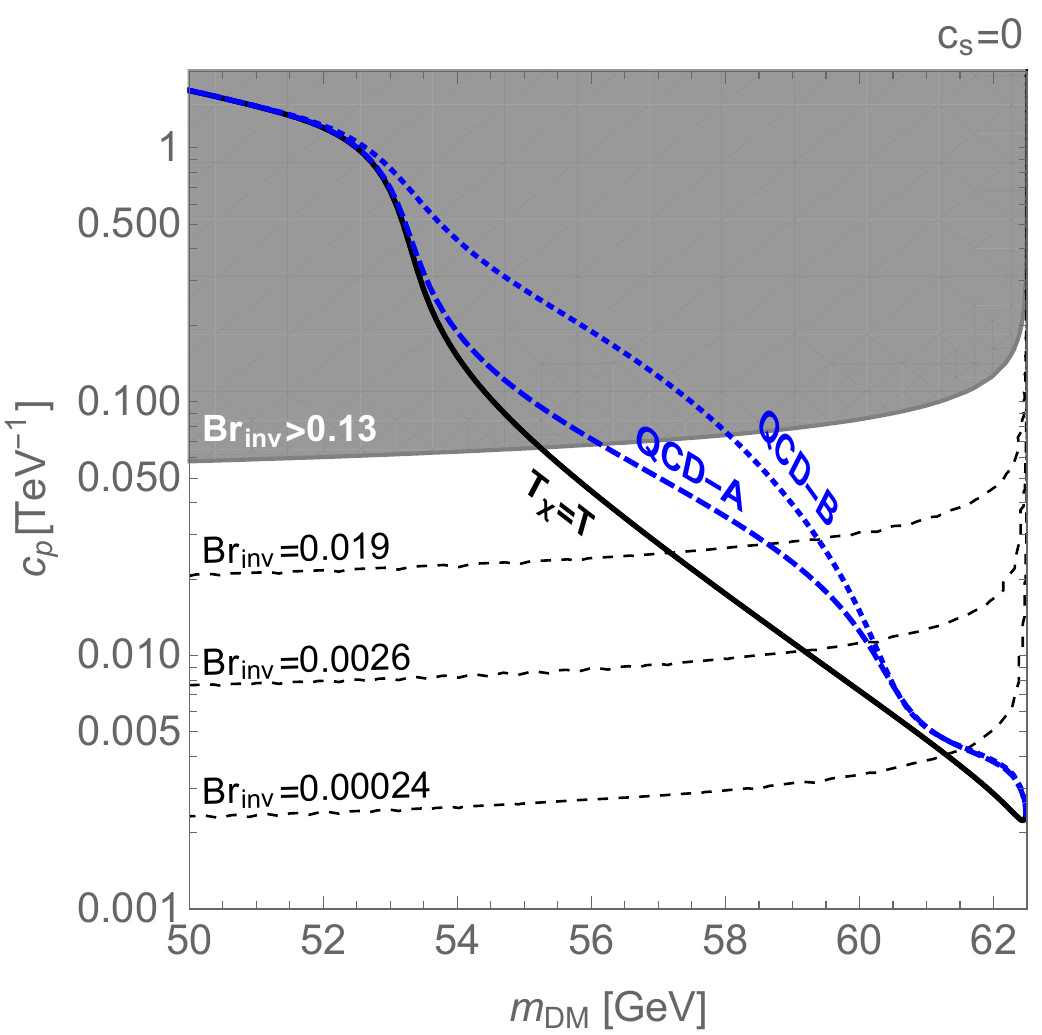}
\quad
\includegraphics[width=0.48\hsize]{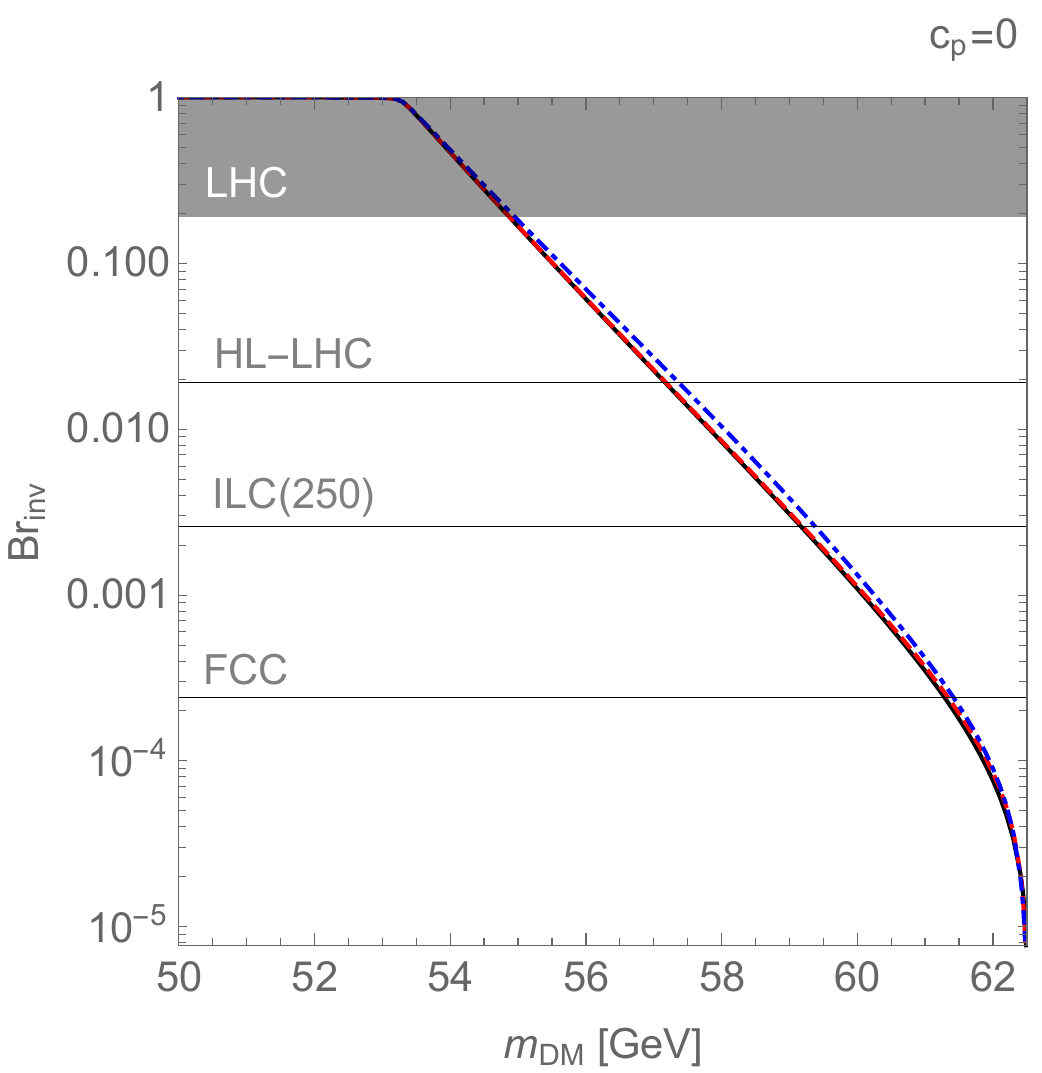}
\caption{
\textbf{Left:}
The values of $c_p$ that explain the measured value of the DM energy density
in the maximally CP-violating case.
The black-solid curve is for the standard calculation without taking into account the effect of the early kinetic decoupling.
The blue-dashed and blue-dotted curves are the results with the effect of the early kinetic decoupling
in the QCD-A and QCD-B scenario, respectively.
The constraint and prospects from the Higgs invisible decay search are also shown.
The gray shaded region is already excluded by the ATLAS and CMS experiments.
The black dashed curves show the prospects of the HL-LHC, ILC, and FCC experiments.
\textbf{Right:}
The branching ratio of the Higgs invisible decay for $c_s = 0$.
The color notations are the same as in the left panel.
}
\label{fig:omegah2}
\end{figure}

We investigate the effect of the early kinetic decoupling in the case for $c_s = 0$, where the CP is maximally violating.
The left panel in Fig.~\ref{fig:omegah2} shows the values of $c_p$ that
explain the measured value of the DM energy density in three scenarios:
the standard calculation ($T_\chi = T$), the QCD-A, and the QCD-B.
We find a significant effect of the early kinetic decoupling.
The larger value of the coupling is required to explain the DM energy density compared to the result with the standard calculation.
In particular, the QCD-B scenario requires at most $\sim 4.5$ times larger coupling.
Even in the QCD-A scenario, which is a conservative scenario for the early kinetic decoupling,
we can see the significant enhancement of $c_p$.
Since the QCD-A and QCD-B are extreme scenarios,
it is expected that the true value of $c_p$ is in between the two curves for the QCD-A and QCD-B in Fig.~\ref{fig:omegah2}.

In the mass range of the DM we are investigating,
the Higgs boson decays into two DM particles.
Since the DM cannot be directly detected at the collider experiments,
this process is known as the Higgs invisible decay.
The larger coupling of the DM to the Higgs boson predicts the larger branching ratio of the Higgs invisible decay.
Since the invisible decay of the Higgs boson is negligible in the SM, 
the large invisible branching ratio is a smoking gun of physics beyond the SM and
is being searched by the ATLAS and CMS experiments. 
Currently, the ATLAS and CMS experiments obtain the upper bound on it as 
\begin{align}
 \text{BR}_\text{inv} <  
\begin{cases}
 0.13 & \text{(ATLAS \cite{ATLAS:2020cjb})}\\
 0.19 & \text{(CMS \cite{1809.05937})} 
\end{cases}
\end{align}
at 95\% CL.
The prospects of various experiments are summarized in \cite{1905.03764},
\begin{align}
 \text{BR}_\text{inv} <
\begin{cases}
 0.019 & \text{(HL-LHC)} \\
 0.0026 & \text{(ILC(250))} \\
 0.00024 & \text{(FCC)}
\end{cases}
\end{align}
at 95\% CL,
where FCC corresponds to the combined performance of 
FCC-ee$_\text{240}$,
FCC-ee$_\text{365}$,
FCC-eh, and FCC-hh.
The prospects for the ILC, and FCC are obtained by combining with the HL-LHC. 
We show the model prediction of the branching ratio of the Higgs invisible decay in the right panel in Fig.~\ref{fig:omegah2} with these prospects and the current bound. 
Due to the large enhancement of $c_p$ by the early kinetic decoupling,
the bound on the mass of the DM is stringent. 
The current lower mass bound on the DM is obtained as $58.1$~GeV in QCD-B,
while it is 55.2~GeV in the standard treatment where the effect of the kinetic decoupling is ignored. The constraint and prospects are also shown in the left panel in Fig.~\ref{fig:omegah2}.

\subsection{With the CP-conserving coupling}

We turn on the CP-conserving coupling $c_s$ and discuss its effect on the kinetic decoupling.
As shown in Eq.~\eqref{eq:amp_DM-f-ela-scatt.}, 
the contribution of the CP-conserving coupling to the DM-quark scattering processes is larger than the contribution of the CP-violating coupling for small $t$.
Therefore, it is expected that the effect of the early kinetic decoupling
is milder for the larger value of $|c_s|$.
We start by investigating whether this expectation is true or not.

\begin{figure}[tb]
\includegraphics[width=0.48\hsize]{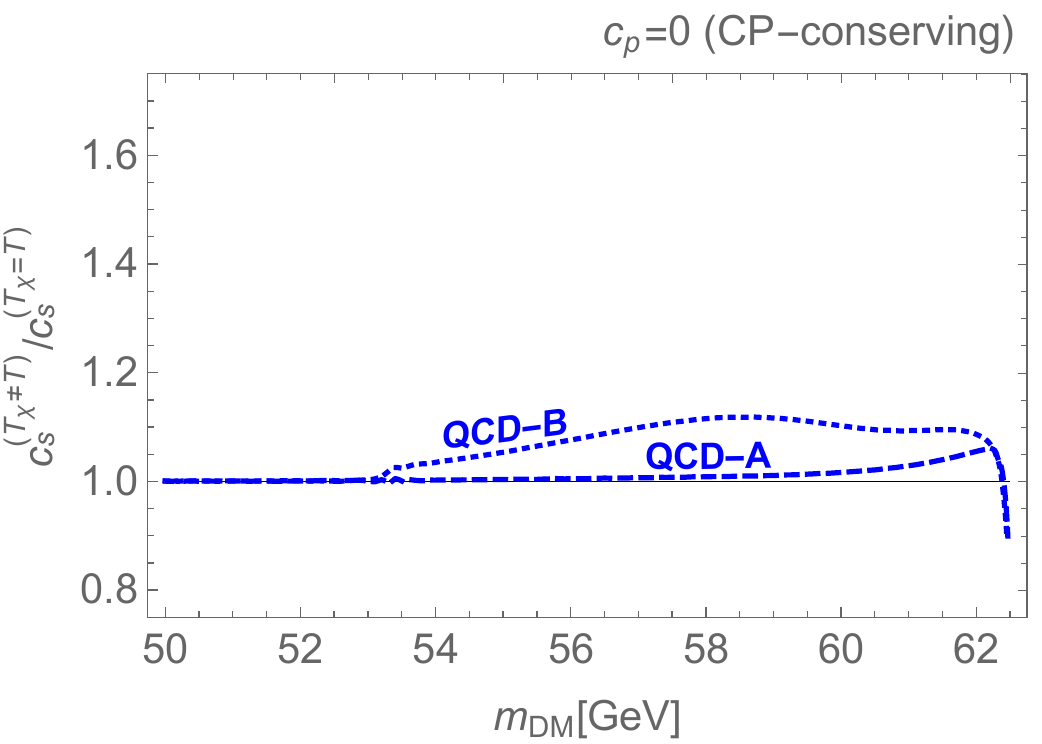} \hspace{4truemm}
\includegraphics[width=0.48\hsize]{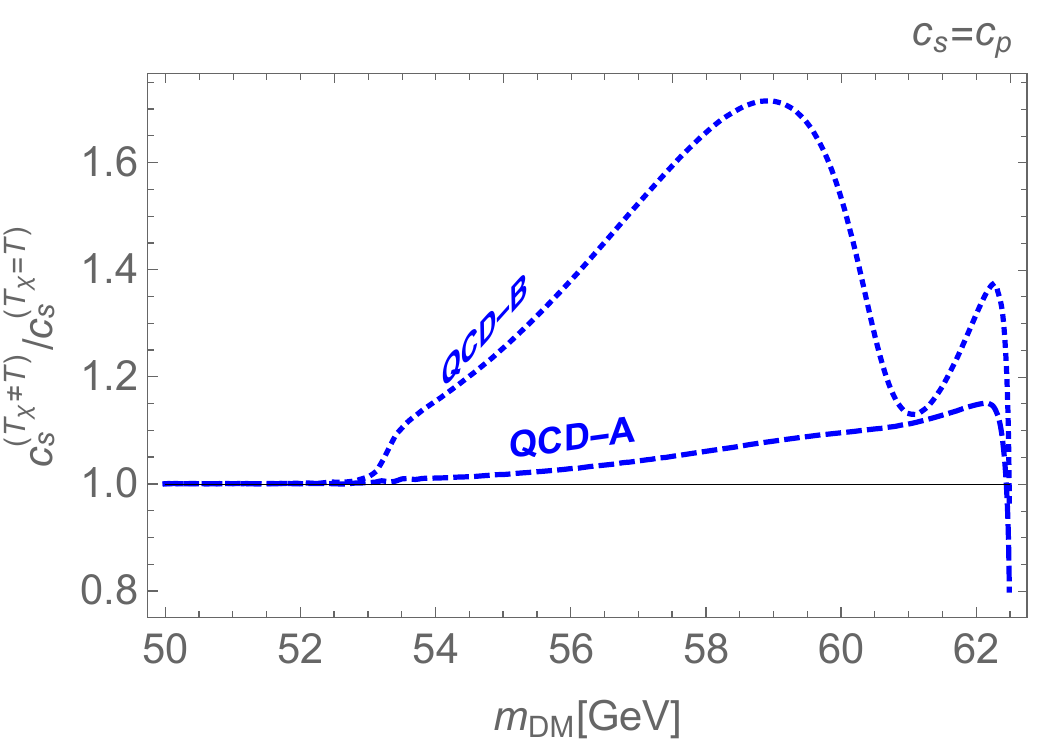}
\includegraphics[width=0.48\hsize]{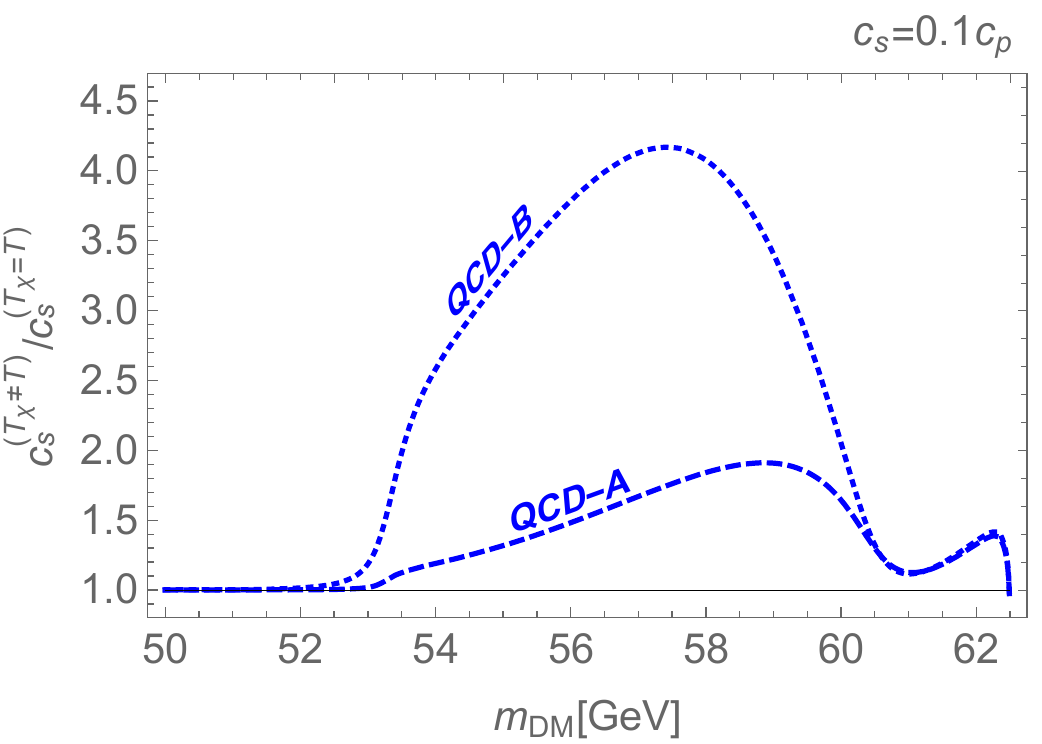} \hspace{4truemm}
\includegraphics[width=0.48\hsize]{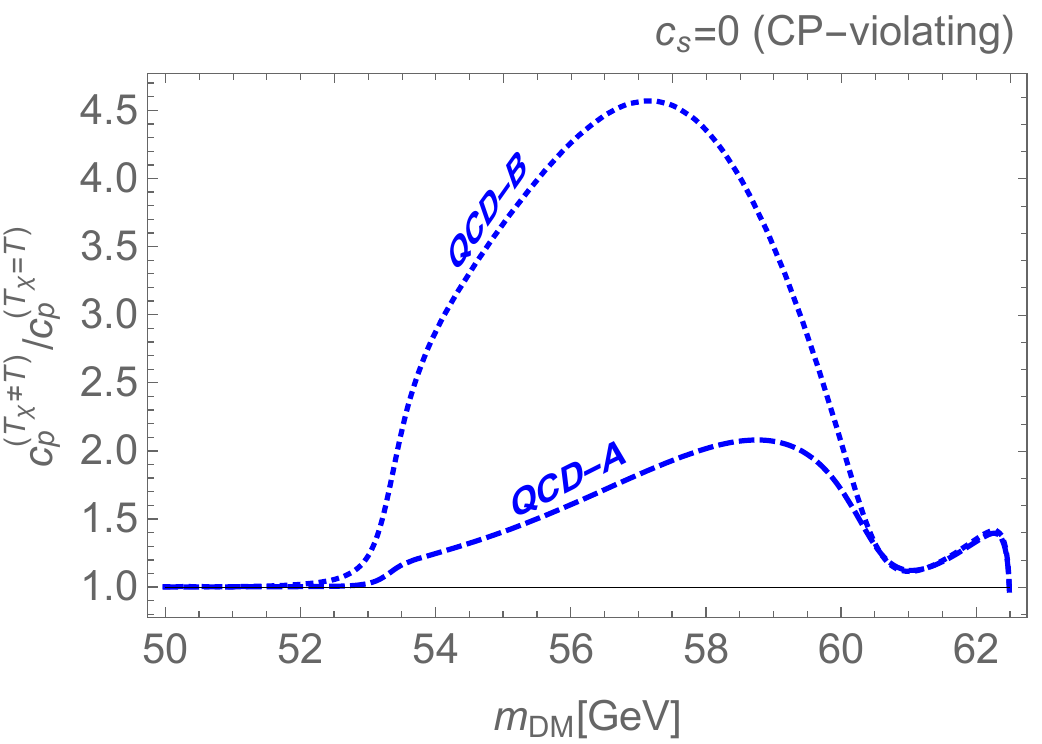}
\caption{
The ratio of the couplings determined with and without the effect of the early kinetic decoupling.
The dashed and dotted curves are the results for the QCD-A and QCD-B scenario, respectively.
}
\label{fig:coupling-enhancement}
\end{figure}

Figure~\ref{fig:coupling-enhancement} shows that the ratio of the couplings determined with and without the effect of the early kinetic decoupling.
The top-left panel shows the case for the maximal CP-conserving case, namely $c_p =0$.
The bottom-right panel is for the maximal CP-violating case that is studied in Sec.~\ref{sec:CPV}.
The other panels are for the mixed cases for some fixed values of the ratio of $c_s$ and $c_p$.
We choose $c_s$ and $c_p$ to obtain the measured value of the DM energy density. 
We find that 
the effect of the early kinetic decoupling is significant once we turn on the CP-violating coupling.
In particular, the coupling enhancement by the early kinetic decoupling in the CP-conserving case is negligible compared to the enhancement in the CP-violating case.
This result is what we expect from
Eqs.~\eqref{eq:amp_DM-f-ela-scatt.} and \eqref{eq:gamma-tilde-F1}. 
We conclude that
if the elastic scattering processes are suppressed by the small momentum transfer,
then the effect of the kinetic decoupling is significant.
It is also found that 
the effect of the early kinetic decoupling is visible even if we have CP-conserving coupling as long as $|c_s| \lesssim 0.1 |c_p|$.

A consequence of the coupling enhancement is the enhancement of the Higgs invisible decay
as discussed in Sec.~\ref{sec:CPV}.
Another consequence with the CP-conserving coupling is the DM-nucleon scattering.
In non-relativistic DM-nucleon scattering processes, only the CP-conserving operator is relevant.
Therefore, the spin-independent cross section ($\sigma_\text{SI}$) is proportional to $c_S^2$ and is given by~\cite{1901.02278},
\begin{align}
 \sigma_{\text{SI}}
~=~
\frac{1}{\pi}
\frac{f_N^2 c_S^2}{m_h^4}
\frac{m_N^4 m_{\chi}^2}{(m_N + m_{\chi})^2}
,
\end{align}
where
\begin{align}
 m_N =& 0.938 \text{ GeV},\\
 f_N =& \frac{2}{9} + \frac{7}{9} \sum_q f_q,\\
 f_u = 0.0110,\ \
 f_d =& 0.0273,\ \
 f_s = 0.0447.
\end{align}
The values of $f_q$ are taken from \texttt{micrOMEGAs}~\cite{1407.6129}.
With the CP-conserving coupling, the combination of Higgs invisible decay searches and the DM direct detection experiments is essential to test the model.

\begin{figure}[tb]
\includegraphics[width=0.48\hsize]{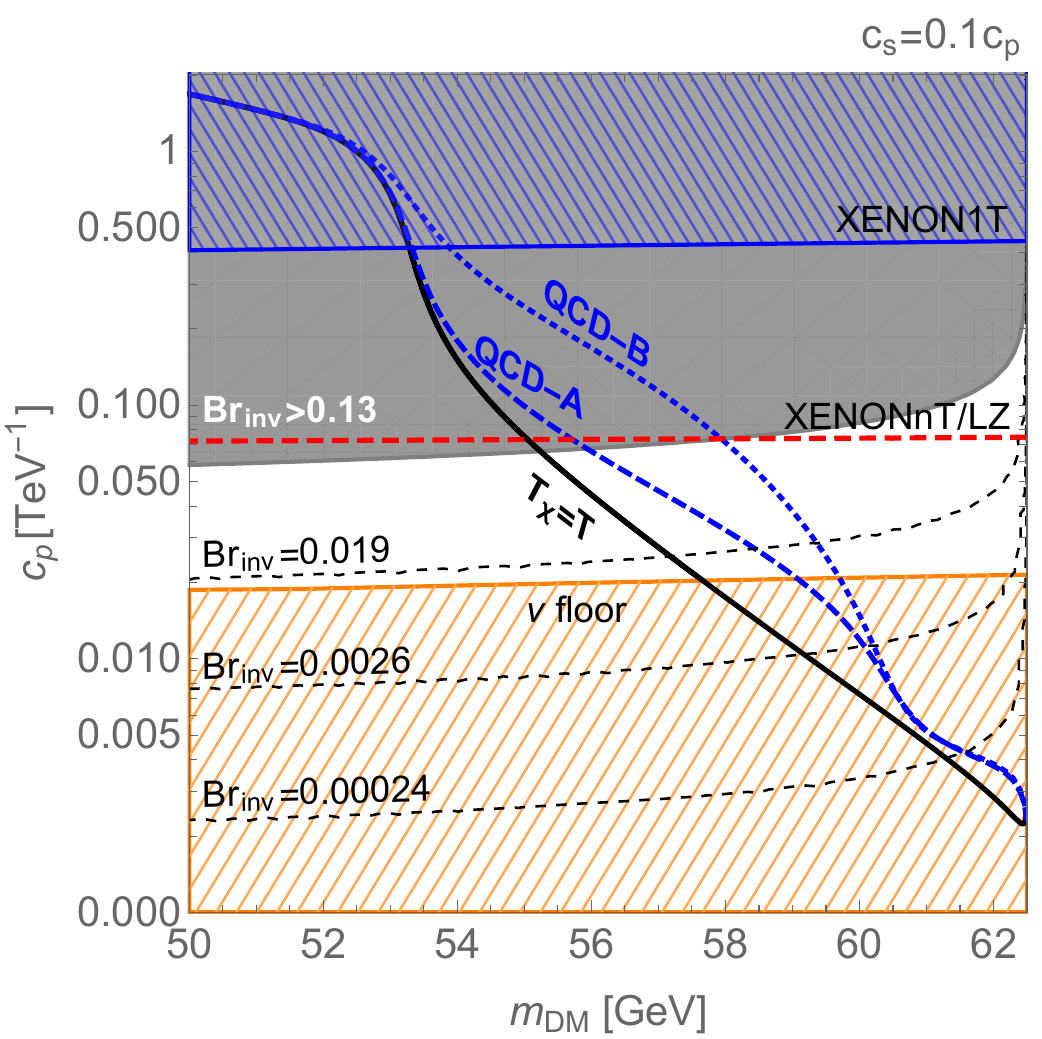}
\includegraphics[width=0.48\hsize]{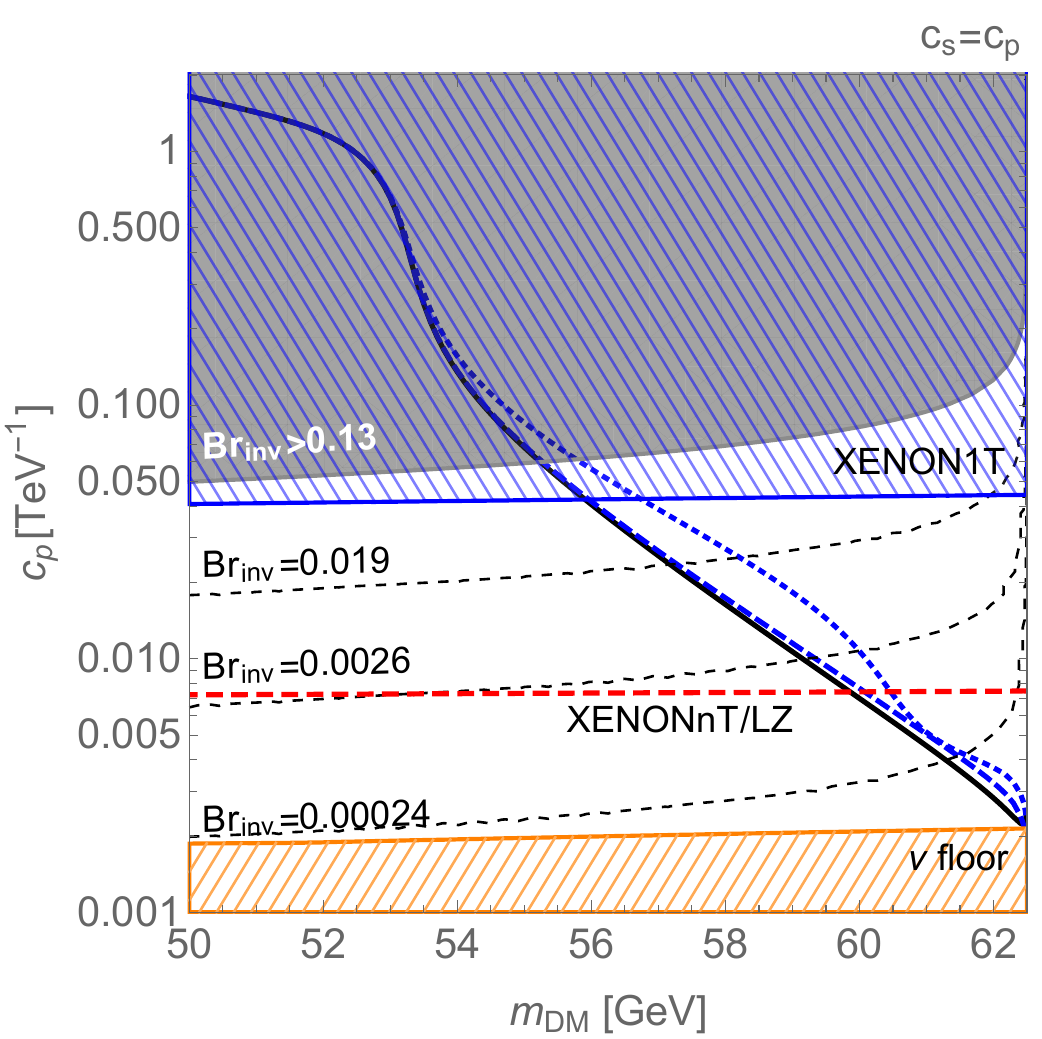}
\includegraphics[width=0.48\hsize]{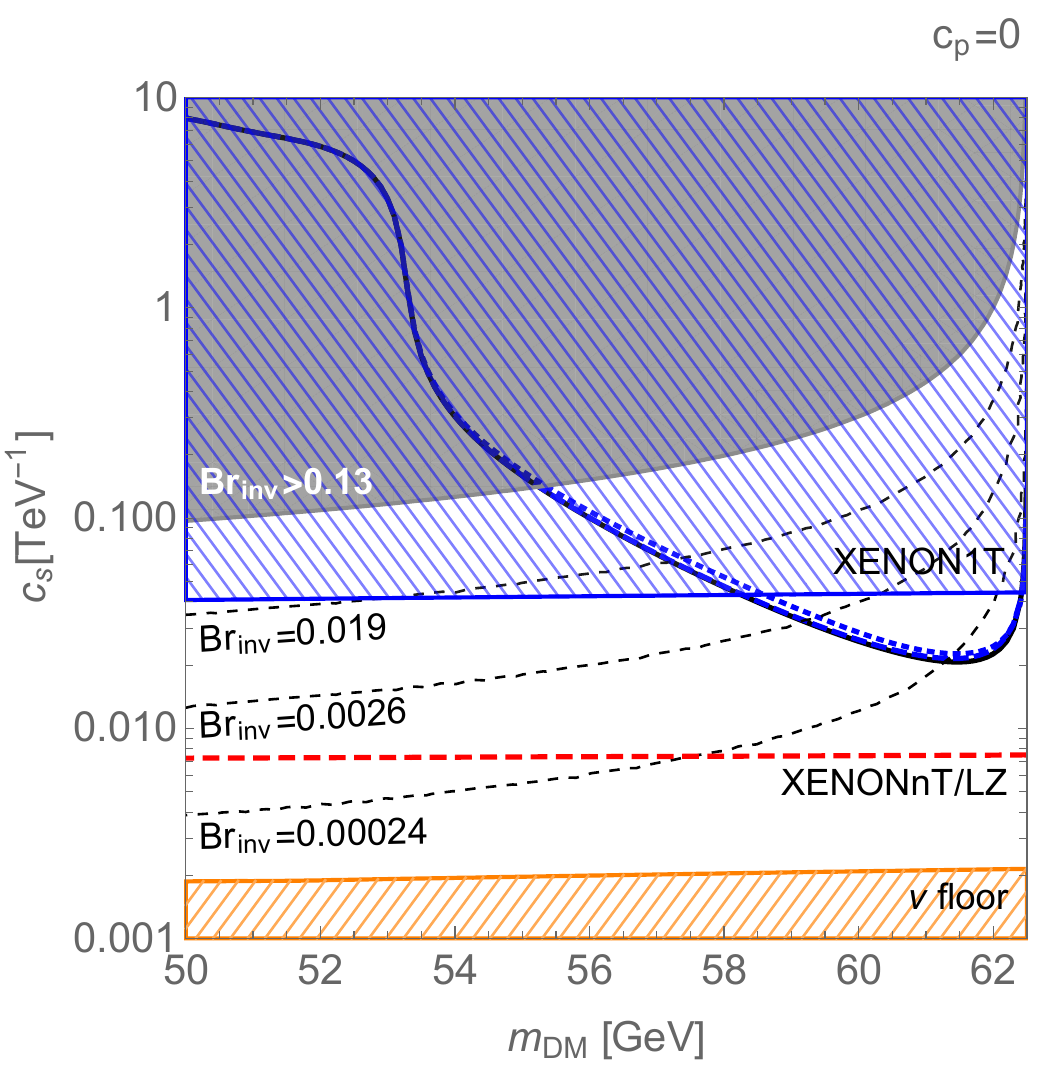}
\caption{
The values of the couplings that explain the measured value of the DM energy density.
The blue-hatched region (\textbackslash\textbackslash\textbackslash) is excluded by the XENON1T experiment~\cite{1805.12562}.
The red-dashed line shows the prospect of the XENONnT and LZ experiments~\cite{Aprile:2015uzo, 1802.06039}.
The orange-hatched region (///) is below the neutrino floor and cannot be accessed by the direct detection experiments. 
The other color notation is the same as in Fig.~\ref{fig:omegah2}.
}
\label{fig:result_w/_CPC}
\end{figure}

Figure~\ref{fig:result_w/_CPC} show the values of $c_p$ (or $c_s$) 
with the constraints and prospects of the Higgs invisible decay and the DM direct detection.
The values of the couplings are determined to reproduce the measured value of the DM energy density.
We find that the constraint from the XENON1T experiment gives a stronger upper bound on the coupling than
the constraint on the Higgs invisible decay for $|c_s| \gtrsim |c_p|$.
For smaller $|c_s|$, the Higgs invisible decay gives the stronger bound on the couplings.
From both constraints, we find that the current lower bound on $m_\chi$ is 55~GeV $\lesssim m_\chi  \lesssim 58$~GeV.
Prospects of the model highly depend on the model parameters.
If the XENONnT or LZ experiments find DM signals, 
then $|c_s|$ is likely to be larger than $\sim 0.1 |c_p|$.
In that case, the Higgs invisible decay can also be detected depending on the mass of DM.
If the XENONnT or LZ experiments observe null results and give upper bounds on $\sigma_\text{SI}$, 
then the maximal CP-conserving case is excluded, and the ILC cannot observe the Higgs invisible decay for $|c_s| \gtrsim |c_p|$. 
For $|c_s| \lesssim 0.1 |c_p|$, null results in the direct detection experiments are consistent, and the search for the Higgs invisible decay is essential to test the model. 

\subsection{Comment on the Quantum correction}

We make some comments on the loop induced diagrams.
Even if we set $c_s = 0$ at the tree level, 
loop diagrams induce the CP-conserving operator.
We estimate the value of $c_s$ induced at the loop level and show that our analysis above does not suffer from the quantum corrections for $|c_s| \ll |c_p|$.

We can estimate the value of $c_s$ induced at the loop level, which we denote as $c_s^\text{loop}$, as follows.
For $|c_s| \ll |c_p|$, $c_s^\text{loop}$ is proportional to $c_p^2$ because
we need to use the CP-violating interaction twice to cancel $\gamma^5$.
$c_s^\text{loop}$ should also be proportional to $m_\chi$, 
because the CP-conserving operator violates the chiral symmetry for the DM, and $m_\chi$ is a source of the breaking of the chiral symmetry in the dark sector.
Therefore, we can estimate $c_s^\text{loop}$ as
\begin{align}
 c_s^\text{loop} \sim \frac{m_\chi c_p^2}{(4\pi)^2}
\simeq
3 \times 10^{-4} 
\left( \frac{m_\chi}{50\text{ GeV}} \right)
\left( \frac{c_p}{\text{TeV}^{-1}} \right)
c_p
.
\end{align}
As shown in Figs.~\ref{fig:omegah2} and \ref{fig:result_w/_CPC},
$|c_p| \ll 1$~TeV$^{-1}$ and thus $|c_p^\text{loop}/c_s| \ll 0.1$.
Since the CP-conserving interaction is negligible for the early kinetic decoupling
if $|c_p^\text{loop}/c_s| \lesssim 0.1$ as can be seen from Fig.~\ref{fig:result_w/_CPC},
we can safely neglect the loop correction to our analysis.

The model we discuss in this paper is non-renormalizable,
and thus $c_s^\text{loop}$ generally depends on the cutoff scale~\cite{1902.11070}. 
We ignored it to estimate the loop effect here.
Since the cutoff scale is a free parameter, our estimation above has ambiguity.
To avoid the ambiguity due to the cutoff scale, 
we have to go beyond the effective theory and work in a UV complete model
such as
the singlet-doublet fermion DM model~\cite{hep-ph/0510064, 0705.4493, 0706.0918}.
CP-odd scalar mediator models~\cite{1404.3716, 1408.4929,1701.04131} are other UV completions.

\section{Conclusion}
\label{sec:summary}

We investigated the effect of the early kinetic decoupling in a model of fermionic DM
that interacts with the SM particles by exchanging of the Higgs boson.
The model has two types of the DM-Higgs couplings, namely the CP-conserving and CP-violating couplings ($c_s$ and $c_p$).
We focused on the DM mass range for 50~GeV $\lesssim m_\chi <$ 62.5~GeV, where
pairs of DM particles annihilate into SM particles efficiently through the Higgs resonance,
and thus the DM-Higgs coupling should be small to obtain the measured value of
the DM energy density by the freeze-out mechanism.
In addition, the elastic scattering processes are suppressed by the small momentum transfer if the DM-Higgs coupling violates the CP invariance.
Therefore, the elastic scattering can be doubly suppressed,
and the temperature of the DM can differ from the temperature of the thermal bath.

For the maximal CP-violating case ($c_s = 0$), 
the model is free from the DM direct detection searches and only constrained by the Higgs invisible searches. After determining the coupling to obtain the measured value of the DM energy density,
we find that the current lower bound on the mass of DM is 55.2~GeV, 56.1~GeV, and 58.1~GeV for the case without taking into account the effect of the early kinetic decoupling,
the QCD-A, and the QCD-B, respectively. 
We find that the ILC experiment can cover for $m_\chi \lesssim 60$~GeV, 
while it is $m_\chi \lesssim 59$~GeV in the analysis that ignores the effect of the early kinetic decoupling.

The effect of non-zero $c_s$ is also studied.
We showed in Fig.~\ref{fig:coupling-enhancement}
that 
the effect of the early kinetic decoupling is significant for $|c_s| \lesssim {\cal O}(0.1)|c_p|$.
The non-zero $c_s$ induces the spin-independent DM-nucleon scattering,
and thus the searches for the Higgs invisible decay and DM direct detection are complements to each other.
For $c_s = 0.1 c_p$, the significant effect from the early kinetic decoupling as shown
in the top-left panel of Fig.~\ref{fig:result_w/_CPC},
and the current constraint from the DM direct detection experiment is much weaker than 
the one from the Higgs invisible decay search.
If the XENONnT/LZ experiments find DM signals in the near future,
then the model predicts that $|c_s| > 0.1 |c_p|$.
For $c_s = c_p$, the kinetic decoupling is sizable for QCD-B but not for QCD-A.
The direct detection experiments are powerful to test the model for this case, 
and the Higgs invisible decay searches are nice complements to it. 
If the XENONnT and LZ experiments discover the DM, 
then future collider experiments discover the Higgs invisible decay. 
For the maximal CP-conserving case ($c_p = 0$), 
the kinetic decoupling does not affect the determination of the coupling for the relic abundance,
and the XENONnT and LZ experiments can cover all the mass range in the Higgs resonance regime
as shown in the bottom panel of Fig.~\ref{fig:result_w/_CPC}.

Although we focused only on the Higgs invisible decay as an observable that is affected by the early kinetic decoupling, 
the early kinetic decoupling is generally expected to have the impact on other observables as well in models that predict suppressed elastic scattering processes.
For example, the electric dipole moment can be affected in the singlet-doublet DM model that is one of the UV completion of the model we discussed in this paper.
Pseudo Nambu-Goldstone DM models~\cite{1708.02253, Abe:2020iph, Okada:2020zxo, Ahmed:2020hiw}
 also predict suppressed elastic scattering processes, and the early kinetic decoupling potentially has the impact on their phenomenology.

\section*{Acknowledgments}
This work is supported in part by JSPS KAKENHI Grant Numbers 16K17715 and 19H04615.
The author wishes to thank 
Tobias Binder, Torsten Bringmann, Michael Gustafsson, Andrzej Hryczuk, and 
Ayuki Kamada for helpful discussions.


\begin{thebibliography}{99}
\bibitem{Lee:1977ua} 
  B.~W.~Lee and S.~Weinberg,
  Phys.\ Rev.\ Lett.\  {\bf 39}, 165 (1977).
  doi:10.1103/PhysRevLett.39.165


\bibitem{1608.07648} 
  D.~S.~Akerib {\it et al.} [LUX Collaboration],
  Phys.\ Rev.\ Lett.\  {\bf 118}, no. 2, 021303 (2017)
  doi:10.1103/PhysRevLett.118.021303
  [arXiv:1608.07648 [astro-ph.CO]].


\bibitem{1708.06917} 
  X.~Cui {\it et al.} [PandaX-II Collaboration],
  Phys.\ Rev.\ Lett.\  {\bf 119}, no. 18, 181302 (2017)
  doi:10.1103/PhysRevLett.119.181302
  [arXiv:1708.06917 [astro-ph.CO]].


\bibitem{1805.12562} 
  E.~Aprile {\it et al.} [XENON Collaboration],
  Phys.\ Rev.\ Lett.\  {\bf 121}, no. 11, 111302 (2018)
  doi:10.1103/PhysRevLett.121.111302
  [arXiv:1805.12562 [astro-ph.CO]].


\bibitem{Griest:1990kh} 
  K.~Griest and D.~Seckel,
  Phys.\ Rev.\ D {\bf 43}, 3191 (1991).
  doi:10.1103/PhysRevD.43.3191


\bibitem{Gondolo:1990dk} 
  P.~Gondolo and G.~Gelmini,
  Nucl.\ Phys.\ B {\bf 360}, 145 (1991).
  doi:10.1016/0550-3213(91)90438-4


\bibitem{1706.07433} 
  T.~Binder, T.~Bringmann, M.~Gustafsson and A.~Hryczuk,
  Phys.\ Rev.\ D {\bf 96}, no. 11, 115010 (2017)
  doi:10.1103/PhysRevD.96.115010
  [arXiv:1706.07433 [astro-ph.CO]].


\bibitem{Silveira:1985rk} 
  V.~Silveira and A.~Zee,
  Phys.\ Lett.\  {\bf 161B}, 136 (1985).
  doi:10.1016/0370-2693(85)90624-0


\bibitem{McDonald:1993ex} 
  J.~McDonald,
  Phys.\ Rev.\ D {\bf 50}, 3637 (1994)
  doi:10.1103/PhysRevD.50.3637
  [hep-ph/0702143 [HEP-PH]].


\bibitem{Burgess:2000yq} 
  C.~P.~Burgess, M.~Pospelov and T.~ter Veldhuis,
  Nucl.\ Phys.\ B {\bf 619}, 709 (2001)
  doi:10.1016/S0550-3213(01)00513-2
  [hep-ph/0011335].


\bibitem{Kanemura:2010sh} 
  S.~Kanemura, S.~Matsumoto, T.~Nabeshima and N.~Okada,
  Phys.\ Rev.\ D {\bf 82}, 055026 (2010)
  doi:10.1103/PhysRevD.82.055026
  [arXiv:1005.5651 [hep-ph]].


\bibitem{1203.2064} 
  L.~Lopez-Honorez, T.~Schwetz and J.~Zupan,
  Phys.\ Lett.\ B {\bf 716}, 179 (2012)
  doi:10.1016/j.physletb.2012.07.017
  [arXiv:1203.2064 [hep-ph]].


\bibitem{1205.3169} 
  A.~Djouadi, A.~Falkowski, Y.~Mambrini and J.~Quevillon,
  Eur.\ Phys.\ J.\ C {\bf 73}, no. 6, 2455 (2013)
  doi:10.1140/epjc/s10052-013-2455-1
  [arXiv:1205.3169 [hep-ph]].


\bibitem{1309.3561} 
  A.~Greljo, J.~Julio, J.~F.~Kamenik, C.~Smith and J.~Zupan,
  JHEP {\bf 1311}, 190 (2013)
  doi:10.1007/JHEP11(2013)190
  [arXiv:1309.3561 [hep-ph]].


\bibitem{Matsumoto:2014rxa} 
  S.~Matsumoto, S.~Mukhopadhyay and Y.~L.~S.~Tsai,
  JHEP {\bf 1410}, 155 (2014)
  doi:10.1007/JHEP10(2014)155
  [arXiv:1407.1859 [hep-ph]].


\bibitem{1512.06458} 
  A.~Beniwal, F.~Rajec, C.~Savage, P.~Scott, C.~Weniger, M.~White and A.~G.~Williams,
  Phys.\ Rev.\ D {\bf 93}, no. 11, 115016 (2016)
  doi:10.1103/PhysRevD.93.115016
  [arXiv:1512.06458 [hep-ph]].


\bibitem{Matsumoto:2016hbs} 
  S.~Matsumoto, S.~Mukhopadhyay and Y.~L.~S.~Tsai,
  Phys.\ Rev.\ D {\bf 94}, no. 6, 065034 (2016)
  doi:10.1103/PhysRevD.94.065034
  [arXiv:1604.02230 [hep-ph]].


\bibitem{1808.10465} 
  P.~Athron {\it et al.} [GAMBIT Collaboration],
  Eur.\ Phys.\ J.\ C {\bf 79}, no. 1, 38 (2019)
  doi:10.1140/epjc/s10052-018-6513-6
  [arXiv:1808.10465 [hep-ph]].


\bibitem{1712.09873} 
  S.~Baum, M.~Carena, N.~R.~Shah and C.~E.~M.~Wagner,
  JHEP {\bf 1804}, 069 (2018)
  doi:10.1007/JHEP04(2018)069
  [arXiv:1712.09873 [hep-ph]].


\bibitem{1602.07624} 
  T.~Binder, L.~Covi, A.~Kamada, H.~Murayama, T.~Takahashi and N.~Yoshida,
  JCAP {\bf 1611}, 043 (2016)
  doi:10.1088/1475-7516/2016/11/043
  [arXiv:1602.07624 [hep-ph]].


\bibitem{1503.03513} 
  M.~Drees, F.~Hajkarim and E.~R.~Schmitz,
  JCAP {\bf 1506}, 025 (2015)
  doi:10.1088/1475-7516/2015/06/025
  [arXiv:1503.03513 [hep-ph]].


\bibitem{Tanabashi:2018oca} 
  M.~Tanabashi {\it et al.} [Particle Data Group],
  Phys.\ Rev.\ D {\bf 98}, no. 3, 030001 (2018).
  doi:10.1103/PhysRevD.98.030001


\bibitem{1807.06209} 
  N.~Aghanim {\it et al.} [Planck Collaboration],
  arXiv:1807.06209 [astro-ph.CO].


\bibitem{1307.1347} 
  S.~Heinemeyer {\it et al.} [LHC Higgs Cross Section Working Group],
  doi:10.5170/CERN-2013-004
  arXiv:1307.1347 [hep-ph].


\bibitem{1205.1914} 
  P.~Gondolo, J.~Hisano and K.~Kadota,
  Phys.\ Rev.\ D {\bf 86}, 083523 (2012)
  doi:10.1103/PhysRevD.86.083523
  [arXiv:1205.1914 [hep-ph]].


\bibitem{0903.0189} 
  T.~Bringmann,
  New J.\ Phys.\  {\bf 11}, 105027 (2009)
  doi:10.1088/1367-2630/11/10/105027
  [arXiv:0903.0189 [astro-ph.CO]].



\bibitem{ATLAS:2020cjb}
 [ATLAS],
ATLAS-CONF-2020-008.


\bibitem{1809.05937} 
  A.~M.~Sirunyan {\it et al.} [CMS Collaboration],
  Phys.\ Lett.\ B {\bf 793}, 520 (2019)
  doi:10.1016/j.physletb.2019.04.025
  [arXiv:1809.05937 [hep-ex]].


\bibitem{1905.03764} 
  J.~de Blas {\it et al.},
  JHEP {\bf 2001}, 139 (2020)
  doi:10.1007/JHEP01(2020)139
  [arXiv:1905.03764 [hep-ph]].


\bibitem{1901.02278} 
  T.~Abe and R.~Sato,
  Phys.\ Rev.\ D {\bf 99}, no. 3, 035012 (2019)
  doi:10.1103/PhysRevD.99.035012
  [arXiv:1901.02278 [hep-ph]].


\bibitem{1407.6129} 
  G.~Bélanger, F.~Boudjema, A.~Pukhov and A.~Semenov,
  Comput.\ Phys.\ Commun.\  {\bf 192}, 322 (2015)
  doi:10.1016/j.cpc.2015.03.003
  [arXiv:1407.6129 [hep-ph]].


\bibitem{Aprile:2015uzo} 
  E.~Aprile {\it et al.} [XENON Collaboration],
  JCAP {\bf 1604}, 027 (2016)
  doi:10.1088/1475-7516/2016/04/027
  [arXiv:1512.07501 [physics.ins-det]].


\bibitem{1802.06039} 
  D.~S.~Akerib {\it et al.} [LUX-ZEPLIN Collaboration],
  Phys.\ Rev.\ D {\bf 101}, no. 5, 052002 (2020)
  doi:10.1103/PhysRevD.101.052002
  [arXiv:1802.06039 [astro-ph.IM]].


\bibitem{1902.11070} 
  F.~Ertas and F.~Kahlhoefer,
  JHEP {\bf 1906}, 052 (2019)
  doi:10.1007/JHEP06(2019)052
  [arXiv:1902.11070 [hep-ph]].


\bibitem{hep-ph/0510064} 
  R.~Mahbubani and L.~Senatore,
  Phys.\ Rev.\ D {\bf 73}, 043510 (2006)
  doi:10.1103/PhysRevD.73.043510
  [hep-ph/0510064].


\bibitem{0705.4493} 
  F.~D'Eramo,
  Phys.\ Rev.\ D {\bf 76}, 083522 (2007)
  doi:10.1103/PhysRevD.76.083522
  [arXiv:0705.4493 [hep-ph]].


\bibitem{0706.0918} 
  R.~Enberg, P.~J.~Fox, L.~J.~Hall, A.~Y.~Papaioannou and M.~Papucci,
  JHEP {\bf 0711}, 014 (2007)
  doi:10.1088/1126-6708/2007/11/014
  [arXiv:0706.0918 [hep-ph]].


\bibitem{1404.3716} 
  S.~Ipek, D.~McKeen and A.~E.~Nelson,
  Phys.\ Rev.\ D {\bf 90}, no. 5, 055021 (2014)
  doi:10.1103/PhysRevD.90.055021
  [arXiv:1404.3716 [hep-ph]].


\bibitem{1408.4929} 
  K.~Ghorbani,
  JCAP {\bf 1501}, 015 (2015)
  doi:10.1088/1475-7516/2015/01/015
  [arXiv:1408.4929 [hep-ph]].


\bibitem{1701.04131} 
  S.~Baek, P.~Ko and J.~Li,
  Phys.\ Rev.\ D {\bf 95}, no. 7, 075011 (2017)
  doi:10.1103/PhysRevD.95.075011
  [arXiv:1701.04131 [hep-ph]].


\bibitem{1708.02253} 
  C.~Gross, O.~Lebedev and T.~Toma,
  Phys.\ Rev.\ Lett.\  {\bf 119}, no. 19, 191801 (2017)
  doi:10.1103/PhysRevLett.119.191801
  [arXiv:1708.02253 [hep-ph]].


\bibitem{Abe:2020iph} 
  Y.~Abe, T.~Toma and K.~Tsumura,
  arXiv:2001.03954 [hep-ph].


\bibitem{Okada:2020zxo} 
  N.~Okada, D.~Raut and Q.~Shafi,
  arXiv:2001.05910 [hep-ph].


\bibitem{Ahmed:2020hiw} 
  A.~Ahmed, S.~Najjari and C.~B.~Verhaaren,
  arXiv:2003.08947 [hep-ph].

\end{thebibliography}
\end{document}